\documentclass[11pt]{article}
\usepackage{moriond,epsfig}

\newcommand{\mb}{M_{\rm bc}}
\newcommand{\de}{\Delta{E}}
\newcommand{\bp}{B^+}
\newcommand{\bz}{B^0}
\newcommand{\pp}{p\bar{p}}

\newcommand{\ppk}{p\bar{p}K^+}
\newcommand{\pppi}{p\bar{p}\pi^+}
\newcommand{\plpi}{p\bar{\Lambda}\pi^-}
\newcommand{\plpiz}{p\bar{\Lambda}\pi^0}
\newcommand{\plg}{p\bar{\Lambda}\gamma}

\begin{document}
\vspace*{4cm}

\title {Baryonic $B$ Meson Decays}
\author { M.Z.~Wang }
                                
\address {Department of Physics, 
National Taiwan University, \\
Taipei, Taiwan, R.O.C.}

\maketitle
\abstracts {
Recent results on baryonic $B$ decays from the two b-factories, 
BABAR and Belle, are presented. These include studies of
$\bp \to \pppi$, $\bp \to \plg$ and $\bz \to \plpi$; observations of 
$\bp \to \plpiz$,
$B \to \Lambda_c^+ \bar\Lambda_c^- K$, and
$\bp \to \bar{\Xi}^0_c  \Lambda_c^+$; and study of the inclusive $B$ decays
to $\Lambda_c$.
%
}

\section{Introduction}

Following the  pioneering work
done by CLEO~\cite{CLEOfu}, various baryonic $B$ decays with charmed or 
charmless baryons in the final states
have been found recently by the two b-factories, BABAR~\cite{BABAR} and  
Belle~\cite{Belle}. The charmed baryonic decays have much larger
branching fractions due to the dominant Cabibbo favored $ b \to c$ transition.
The charmless modes presumably proceed via the $b \to s$ penguin
or the $b \to u$ tree processes. The charmed baryonic $B$ decays are 
observed in four-body, three-body and two-body final states while there are
only three-body final states being found for the charmless case.
There is a common
feature for the charmless decays that the baryon-antibaryon mass
spectra peak near threshold. This feature was conjectured
in Ref.~\cite{HS} and has recently aroused much
theoretical interest.

In b-factory, it is an over-constrained system to determine decays
from $B$ mesons since not only the mass but also 
the energy of the $B$ meson are known in the center-of-mass (CM)
frame.
One can pick two kinematic variables in the CM frame to identify the
reconstructed $B$ meson candidates, for example, the beam energy
constrained mass $\mb = \sqrt{E^2_{\rm beam}-p^2_B}$, and the
energy difference $\de = E_B - E_{\rm beam}$, where $E_{\rm
beam}$ is the beam energy, and $p_B$ and $E_B$ are the momentum and
energy, respectively, of the reconstructed $B$ meson.
After performing various selection cuts for background suppression,  
the $B$ yields can be determined by an unbinned extended 
likelihood fit using the above two variables as
inputs for all candidate events. The signal probability density function
(PDF) of the two variables is typically obtained by Monte Carlo samples and 
the background PDF is determined from sideband (i.e. non-signal region) data.

\section{Charmless modes}

After the first observation of the charmless baryonic $B$ meson decay,
$\bp \to \ppk$~\cite{belleppk,conjugate}, 
many charmless three-body baryonic decays were
found.  Detailed information from
the polar angle  distributions~\cite{polar} and Dalitz plot~\cite{BABARppk}
offer better understanding of the underlying dynamics. 
We use a data sample consisting of  
$449 \times 10^6 B\bar{B}$ pairs to study
the baryon angular distribution in the proton-antiproton
helicity frame with $M_{\pp} < 2.85$ GeV/$c^2$ for the decays of  
$\bp \to \ppk$ and $\bp \to \pppi$. 
This angle is defined between the baryon direction and 
the oppositely charged meson
direction in the
proton-antiproton pair rest frame.
The observed angular  distributions
for the two modes have opposite trends.
Further theoretical investigations are needed to explain the
behavior of  $\ppk$ and $\pppi$ modes simultaneously.

\begin{figure}[htb]
\begin{center}
\epsfig{file=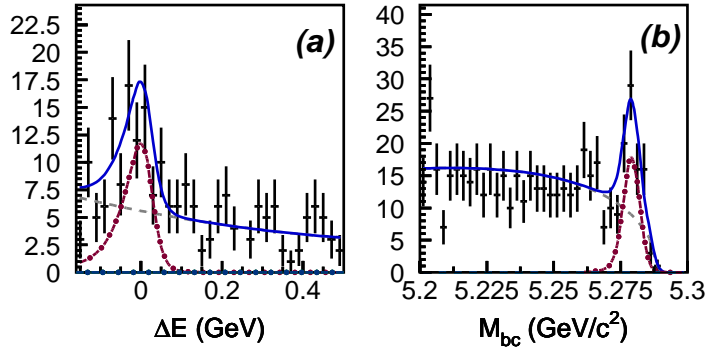, width=4.in}
\caption{ The (a) $\de$ and (b) $\mb$ distributions for
the $\plpiz$ mode with the requirement of
baryon-antibaryon mass $< 2.8$ GeV/$c^2$. The solid curve
represents the fit projection,
which is the sum of signal (dash-dotted peak)
and background (dashed curve) estimations.
}
\label{fg:mergembde}
\end{center}
\end{figure}

We also study the decays of $\bp \to \plg$, $\bp \to \plpiz$ 
and $\bz \to \plpi$.
Note that the results include the
first observation of $\bp \to \plpiz$.
Figure~\ref{fg:mergembde} illustrates the fits for the $B$ yields
in a baryon-antibaryon mass region below 2.8 GeV/$c^2$ for the $\plpiz$ mode.
The ratio of ${\cal{B}}(\bp\to\plpiz) /{\cal{B}}(\bz\to\plpi)$
is $0.93^{+0.21}_{-0.19} \pm 0.09$, which is larger than the theoretical
prediction of 0.5.
We also study the two-body intermediate decays
$\bz \to p {\bar\Sigma}^{*-}$, $\bz \to \Delta^0 \bar{\Lambda}$,
$\bp \to p {\bar\Sigma}^{*0}$, and  $\bp \to \Delta^+ \bar{\Lambda}$, where
the $\bar\Sigma^{*-,*0}$ and $\Delta^{0,+}$ are reconstructed in the
$\bar\Sigma^{*-,*0}\to \bar\Lambda\pi^{-,0}$
and $\Delta^{0,+} \to p \pi^{0,+}$ channels, respectively. The selection
criteria are
$1.30$ GeV/$c^2$ $< M_{\bar\Lambda\pi^{-,0}}<1.45$ GeV/$c^2$ and
$M_{p\pi^{0,+}} < 1.40$ GeV/$c^2$.
No significant signals are found in these decay chains.
We set upper limits
on the branching fractions at the 90\% confidence
level using the methods described in Refs.~\cite{Gary,Conrad}, 
where the systematic uncertainty is
taken into account. The results are listed in Table~\ref{br-results}.

In the low mass region below 2.8 GeV/$c^2$,
we study the proton angular distribution of the baryon-antibaryon pair system.
The angle $\theta_p$ is defined as the
angle between the proton direction and the meson (photon) direction in the
baryon-antibaryon pair rest frame.
We define the angular asymmetry as $A_{\theta} = {
{Br_+ - Br_-}\over
{Br_+ + Br_-}}$, where $Br_+$ and
$Br_-$
stand for the measured branching fractions with $\cos\theta_p > 0$ and
 $\cos\theta_p < 0$, respectively. The measured results are shown in
Table~\ref{br-results}.
We also measure the charge asymmetry
as $A_{CP}$= $(N_{b} - N_{\bar{b}})/ (N_{b} + N_{\bar{b}})$ for these modes,
where $b$ stands for the quark flavor of the $B$ meson. 
The results are listed in Table~\ref{br-results}.
The measured charge asymmetries are consistent with zero within
their statistical uncertainties.

\begin{table}[htb]
\caption{Summary of the results. Y is the fitted signal or upper limit
at 90\% confidence, $\sigma$ is the statistical significance,
{$\cal{B}$} is the branching fraction,
$A_{\theta}$ is the angular asymmetry and
$A_{CP}$ is the charge asymmetry.}
\label{br-results}
\begin{center}
\begin{tabular}{c|ccccc}
Mode & \ Y & $\sigma$ & {$\cal B$} ($10^{-6}$)  & $A_{\theta}$  & $A_{CP}$ \\
\hline $\bp\to\plg$ & \ $114^{+18}_{-16}$ & 14.5 & $2.45^{+0.44}_{-0.38}\pm 0.22$ &
$0.29 \pm 0.14 $ & $0.17\pm 0.17 $
\\
\hline $\bp\to\plpiz$ & \ $89^{+19}_{-17}$ & 10.2 & $3.00^{+0.61}_{-0.53}\pm 0.33$ &
$-0.16 \pm 0.18 $ & $0.01 \pm 0.17$
\\
       ~~~~~~~~~~$\bp \to p {\bar\Sigma}^{*0}$ & \ $<11.3$& - & $<0.47$ & - & -
\\
       ~~~~~~~~~~$\bp \to \Delta^+ \bar{\Lambda}$ & \ $<15.9$ & - &$<0.82$ & - & -
\\
\hline $\bz\to\plpi$ & \ $178^{+18}_{-16}$ & 20.0 & $3.23^{+0.33}_{-0.29}\pm 0.29$ &
$-0.41 \pm 0.11 $ &  $-0.02 \pm 0.10$
\\
       ~~~~~~~~~~$\bz \to p {\bar\Sigma}^{*-}$ & \ $<10.9$ & -& $<0.26$ & - & -
\\
       ~~~~~~~~~$\bz \to \Delta^0 \bar{\Lambda}$ & \ $<15.9$ & -&$<0.93$ & - & -\\
\end{tabular}
\end{center}
\end{table}

\section{Charmed Modes}

The $b \to c$ process is the dominant process for $B$ decays. Many decay modes
with  $\Lambda_c^+$ in the final states have been found, including the
first observation 
of a two-body decay:$\bar{B}^0 \to \Lambda_c^+ \bar{p}$~\cite{Lcp}.
It is interesting to see that
the two-body baryonic decay is suppressed in exclusive $B$ decays in
contrast to  mesonic $B$ decays where two-body and three-body decays are 
comparable. This 
indicates that, for the formation of a baryon-antibaryon pair, 
giving off extra 
energy is much favored.
Similar decay processes via the $b \to c\bar{c}s$ transition and
 limited phase space have been found for 
$\bp \to \Lambda_c^+ \bar\Lambda_c^- K^+$ and $\bz \to \Lambda_c^+ 
\bar\Lambda_c^- K^0$~\cite{LcLcK}
 in a  $386 \times 10^6 B\bar{B}$ data sample.
The measured branching fractions are unexpectedly large: 
${\mathcal B}(\bp \to \Lambda_c^+ \bar\Lambda_c^- K^+) = 
(6.5^{+1.0}_{-0.9} \pm 1.1 \pm 3.4)\times 10^{-4}$ and 
${\mathcal B}(\bz \to \Lambda_c^+ \bar\Lambda_c^- K^0) = 
(7.9^{+2.9}_{-2.3} \pm 1.2 \pm 4.1)\times 10^{-4}$, where the first error
represents the statistical uncertainty, the second error is the 
systematic error and the last error is due to the 52\% uncertainty in the 
absolute branching fraction of
$\Lambda_c^+ \to p K^- \pi^+$.
This large rate might be understood by the threshold enhancement
phenomenon.
Observation of this kind
of decay is important for the determination of the charm particle yield
per $B$ decay. Decays like $B \to \Lambda_c^+ \bar\Lambda_c^- K$ would 
give a wrong-sign $\Lambda_c^+$, where for most cases only  
$\bar\Lambda_c^-$'s are present in the final state from $B$ decays.

\begin{figure}[htb]
\begin{center}
\epsfig{file=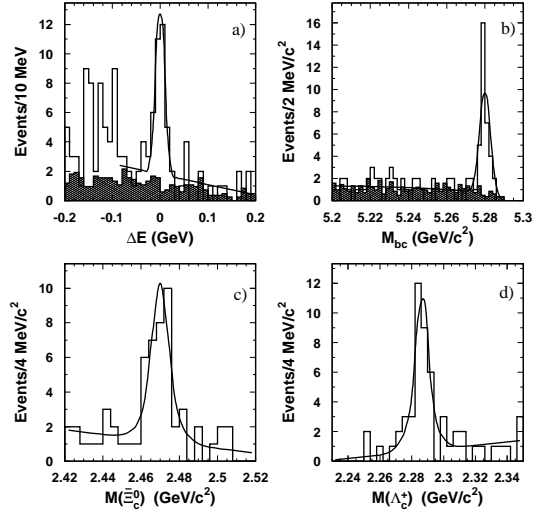, width=3.in}
\caption{ The (a) $\de$ and (b) $\mb$ distributions for
the $\bp \to \bar{\Xi}^0_c  \Lambda_c^+$ candidates. The hatched histograms
show the combined $\bar{\Xi}^0_c$ and $\Lambda_c^+$ mass sidebands 
normalized to the signal region. The excess around low $\de$ region 
maybe be due to
decays with extra final state particle, e.g. $\bar{\Xi}^0_c
\Lambda_c^+ \pi^0$. The (c) $\bar{\Xi}^0_c$ and (d) $\Lambda_c^+$ mass
distributions for candidates taken from the $B$-signal region. The overlaid
curves are the fit results.
}
\label{fgScLc2}
\end{center}
\end{figure}

Another doubly charmed baryonic 
two-body decay,
$\bp \to \bar{\Xi}^0_c  \Lambda_c^+$,
has been found in the same data set. 
Judging from the similarity between  
$ b \to c\bar{c}s$ and $b \to c\bar{u}d$, 
one would
expect these two decay modes, $\bp \to \bar{\Xi}^0_c  \Lambda_c^+$
and $\bar{B}^0 \to \Lambda_c^+ \bar{p}$,
 to have similar branching fractions. However,
the measured  branching fraction product~\cite{XiLc} 
${\mathcal B}(\bp \to \bar{\Xi}^0_c  \Lambda_c^+) \times 
{\mathcal B}(\bar{\Xi}^0_c \to  \bar{\Xi}^+\pi^-) =
(4.8^{+1.0}_{-0.9} \pm 1.1 \pm 1.2)\times 10^{-4}$ is  too big. Assuming
${\mathcal B}(\bar{\Xi}^0_c \to \bar{\Xi}^+\pi^-)$ is at 1\% level, then
${\mathcal B}(\bp \to \bar{\Xi}^0_c  \Lambda_c^+) \sim 10^{-3}$. This is 
about 100
times bigger than that of $\bar{B}^0 \to \Lambda_c^+ \bar{p}$. 
This is another example
of a large enhancement for smaller available energy in 
the baryon-antibaryon system.
Fig.~\ref{fgScLc2} shows the observed signals.

The question of charm counting for $B$ decays is a fundamental issue
to be addressed. Using fully reconstructed $B$ sample (flavor tagging), one
can study the correlated ($b \to c$) and 
anti-correlated ($b \to \bar{c}$) charm production.
Presumably the correlated charm production is dominant and the anti-correlated
is suppressed with a phase space factor. Experimentally, one can determine
the correlated/anti-correlated charm production by studying the inclusive
$B$ decay rates to a limited charm hadrons, e.g. $\Lambda_c^+$, $D^0$, etc.,
because
all other heavier charm particles decay into one of these special cases. 
Recently, BABAR used a $231  \times
10^6 B\bar{B}$ event sample and made the measurement~\cite{taggedLc}.  The 
charm yield per $B$ decay is around 1.2.

\section{Summary}
The exclusive baryonic $B$ decays are well established 
after a few years running
of the two b-factories. One important thing for baryonic $B$ meson decays
is to understand the threshold enhancement mechanism. The study 
of baryonic $B$ decays
is booming with rapidly accumulating data samples. Many new results including
CP violation measurements can be expected in the very near future.

\section*{Acknowledgments}
The author wishes to thank the Moriond QCD organization committee
for making such a wonderful conference. This work is supported by the
National Science Council of the Republic of China under the grant
NSC-95-2119-M-002-033.


\begin{thebibliography}{99}
\bibitem{conjugate} {Throughout this report,
inclusion of charge conjugate mode is always implied unless
otherwise stated.}
\end{thebibliography}
\end{document}